\documentclass{sf2a-conf2014}
\usepackage{graphicx}
\usepackage{hyperref}
\usepackage[]{natbib}  
\usepackage{epstopdf}

\def\BibTeX{{\rm B\kern-.05em{\sc i\kern-.025em b}\kern-.08em
    T\kern-.1667em\lower.7ex\hbox{E}\kern-.125emX}}
\bibpunct{(}{)}{;}{a}{}{,}  

\begin{document}

\TitreGlobal{SF2A 2014}


\title{The BRITE spectropolarimetric survey}

\runningtitle{The BRITE spectropolarimetric survey}

\author{C. Neiner}\address{LESIA, Observatoire de Paris, CNRS UMR 8109, UPMC,
Universit\'e Paris Diderot, 5 place Jules Janssen, 92190 Meudon, France}

\author{A. L\`ebre}\address{LUPM - UMR 5299 - CNRS and Universit\'e Montpellier II - place E. Bataillon, 34090 Montpellier, France}

\setcounter{page}{237}


\maketitle


\begin{abstract}
The BRITE constellation of nanosatellites observes very bright stars to perform
seismology. We have set up a spectropolarimetric survey of all BRITE targets,
i.e. all ~600 stars brighter than V=4, with Narval at TBL, ESPaDOnS at CFHT and
HarpsPol at ESO. We plan to reach a magnetic detection threshold of B$_{\rm pol}
= 50$ G for stars hotter than F5 and B$_{\rm pol} = 5$ G for cooler stars. This
program will allow us to combine magnetic information with the BRITE seismic
information and obtain a better interpretation and modelling of the internal
structure of the stars. It will also lead to new discoveries of very bright
magnetic stars, which are unique targets for follow-up and multi-technique
studies.
\end{abstract}



\section{BRITE}

The BRITE (BRIght Target Explorer) constellation of nano-satellites monitors
photometrically, in 2 colours, the brightness and temperature variations of
stars with V$\le$4, with high precision and cadence, in order to perform
asteroseismology \citep{weiss2014}. The mission consists of 3 pairs of
nano-satellites, built by Austria, Canada and Poland, carrying 3-cm aperture
telescopes. One instrument per pair is equipped with a blue filter; the other
with a red filter. Each BRITE instrument has a wide field of view
($\sim$24$^\circ$), so up to 25 bright stars can be observed simultaneously, as
well as additional fainter targets with reduced precision. Each field will be
observed during several months. As of September 2014, 6 nano-satellites are
already flying and 5 are observing. Each pair of nano-satellites can (but does
not have to) observe the same field and thus increase the duty cycle of
observations. 

BRITE primarily measures pressure and gravity modes of pulsations to probe
the interiors and evolution of stars through asteroseismology. Since the BRITE
sample consists of the brightest stars, it is dominated by the most
intrinsically luminous stars: massive stars at all evolutionary stages, and
evolved cooler stars at the very end of their nuclear burning phases (cool
giants and AGB stars). Analysis of OB star variability will help solve two
outstanding problems: the sizes of convective (mixed) cores in massive stars and
the influence of rapid rotation on their structure and evolution. In addition,
measurements of the timescales involved in surface granulation and differential
rotation in AGB stars, cool giants and cool supergiants will constrain turbulent
convection models.

\begin{figure}[ht!]
 \centering
 \includegraphics[width=0.36\textwidth,clip]{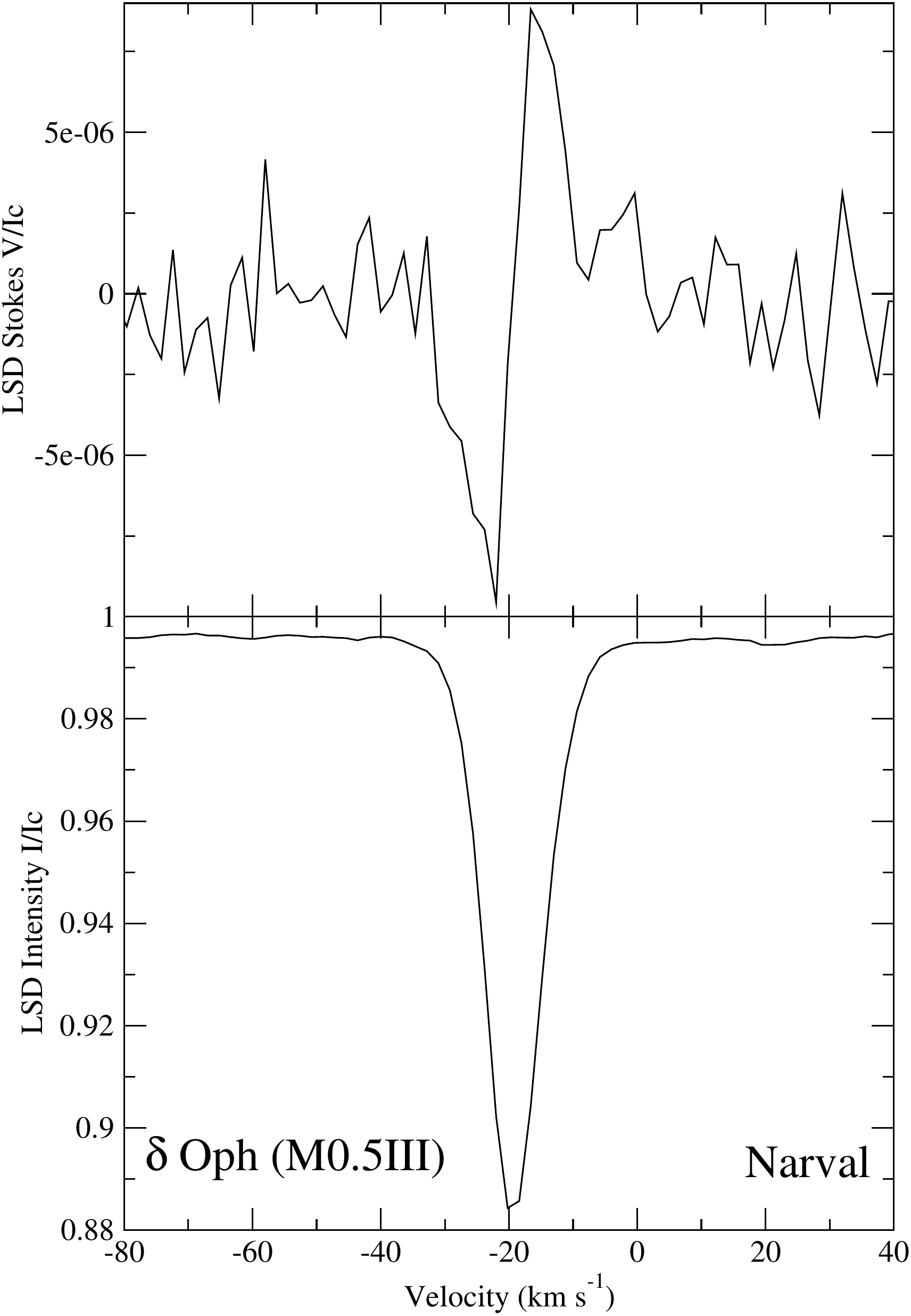}%
 \includegraphics[width=0.36\textwidth,clip]{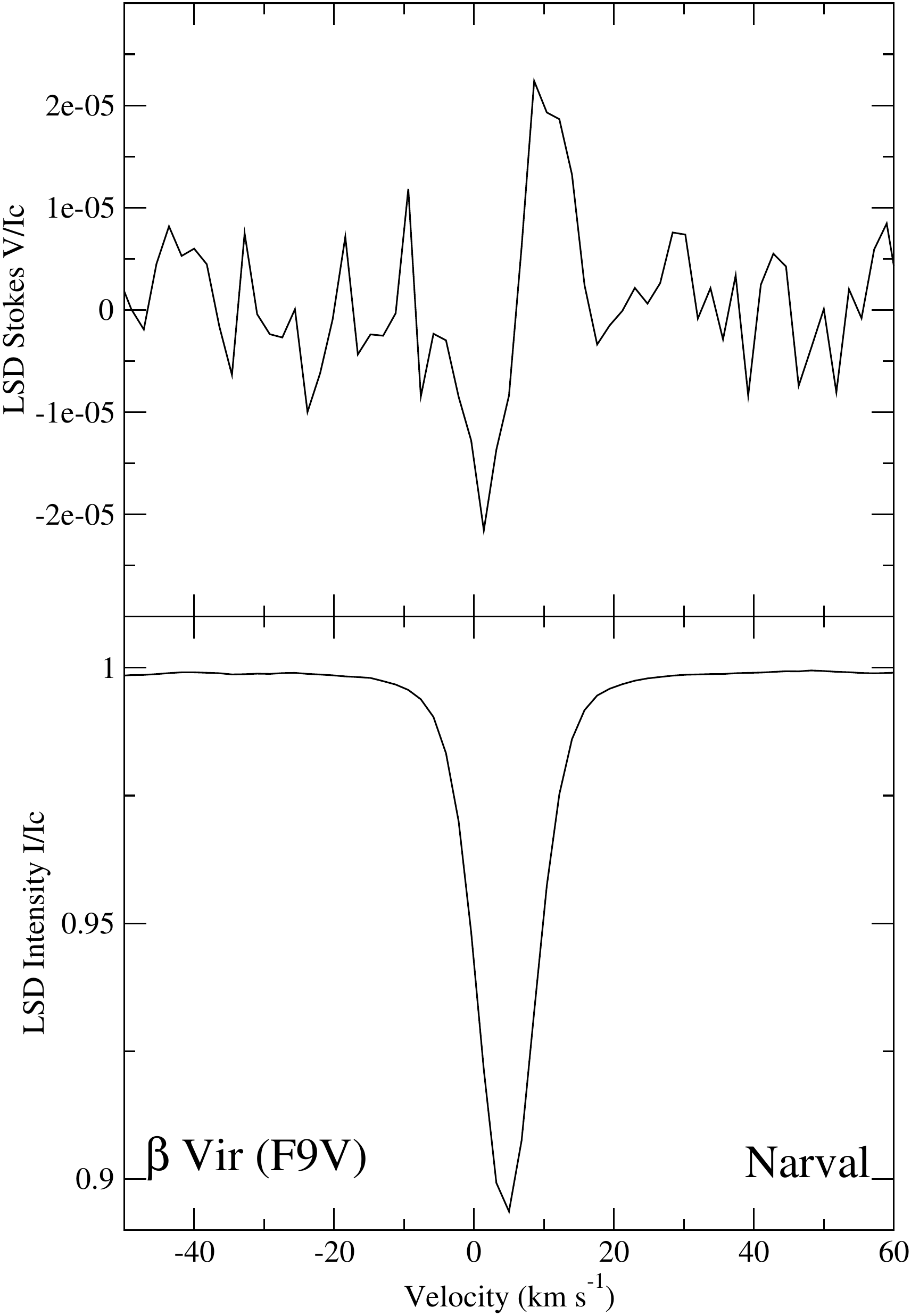}      
  \caption{Examples of magnetic field detections in the single cool stars
$\delta$\,Oph and $\beta$\,Vir. Shown are the LSD Stokes V magnetic signatures
(top panels) and LSD intensity profiles (bottom panels).}
  \label{singlecool}
\end{figure}

\section{Combining asteroseismology and spectropolarimetry}

The study of the magnetic properties of pulsating stars is particularly
interesting since, when combined with the study of their pulsational properties,
it provides (1) a unique way to probe the impact of magnetism on the physics of
non-standard mixing processes inside these stars and (2) strong constraints on
seismic models thanks to the impact of the field on mode splittings and
amplitudes.

The combination of an asteroseismic study with a spectropolarimetric study has
been accomplished for only a couple of massive stars so far, e.g. for the
$\beta$ Cep star V2052\,Oph \citep{briquet2012}. This star presents a magnetic
field with a strength at the poles of about 400 G that has been modelled thanks
to Narval spectropolarimetry \citep{neiner2012}. Moreover our asteroseismic
investigations of this object showed that the stellar models explaining the
observed pulsational behaviour do not have any convective core overshooting
\citep{briquet2012}. This outcome is opposite to other results of dedicated
asteroseismic studies of non-magnetic $\beta$~Cep stars
\citep[e.g.][]{briquet2007}. Indeed, it is usually found that convective core
overshooting needs to be included in the stellar models in order to account for
the observations \citep{aerts2010}. The most plausible explanation is that the
magnetic field inhibits non-standard mixing processes inside V2052\,Oph. Indeed
the field strength observed in V2052\,Oph is above the critical field limit
needed to inhibit mixing determined from theory \citep[e.g.][]{zahn2011}. These 
findings opened the way to a reliable exploration of the effects of magnetism on
the physics of mixing inside stellar interiors of main-sequence B-type
pulsators. 

Conversely, the deformation of line profiles by pulsations is usually neglected
when modelling the magnetic field present in pulsating stars. However, these
deformations directly impact the shape of the Stokes V signatures and thus our
ability to derive correct magnetic parameters. We recently developed a version
of the Phoebe 2.0 code that allows us to model both the line and Stokes V
profiles at the same time, taking pulsations into account, thus presenting for
the first time coherent spectropolarimetric models including magnetism and
pulsations \citep[see][]{neiner2014}. Thanks to this work, and the combination
of seismic and spectropolarimetric data, much more reliable magnetic parameters
can be derived for pulsators.

\begin{figure}[ht!]
 \centering
 \includegraphics[width=0.4\textwidth,clip]{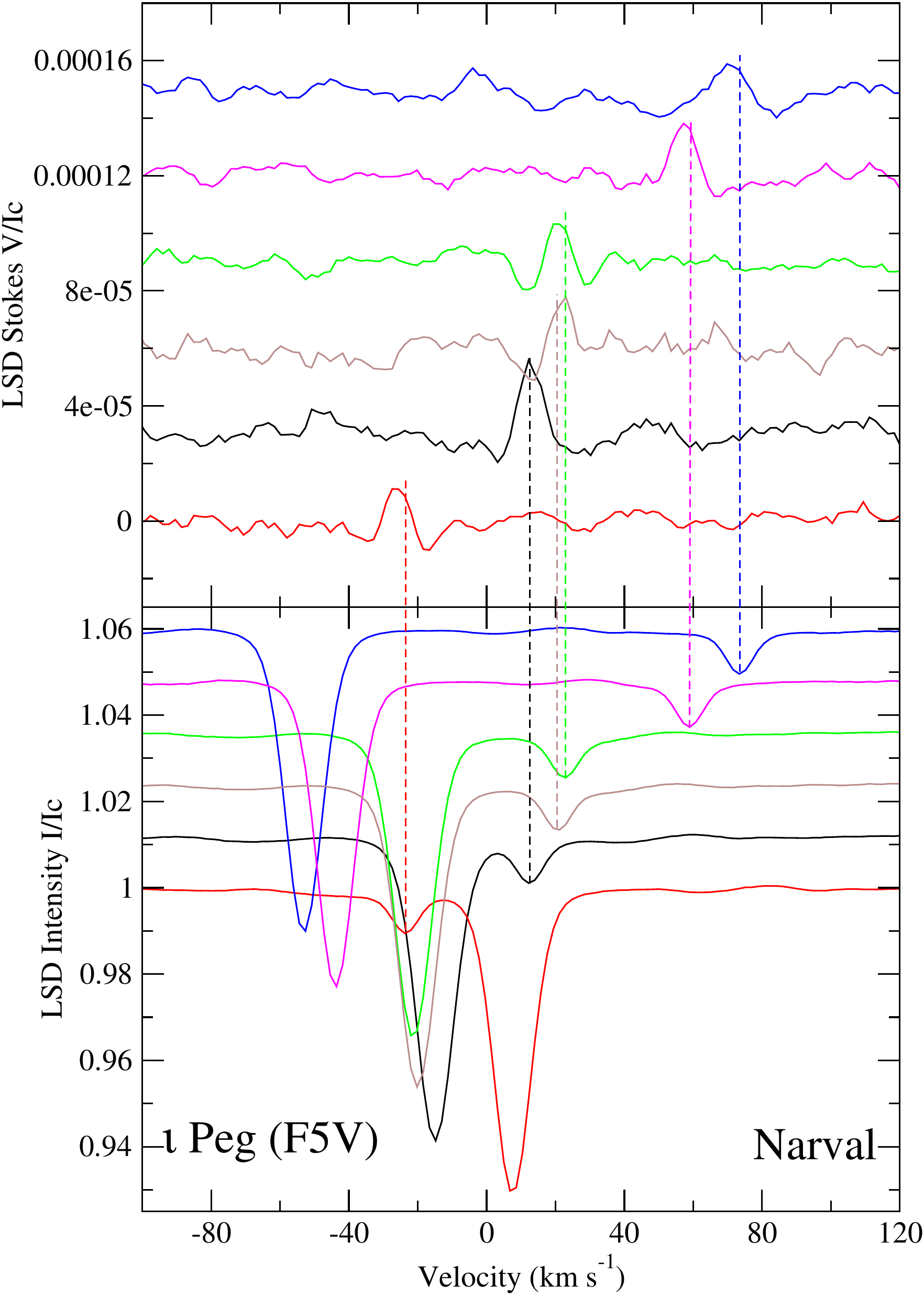}%
 \includegraphics[width=0.4\textwidth,clip]{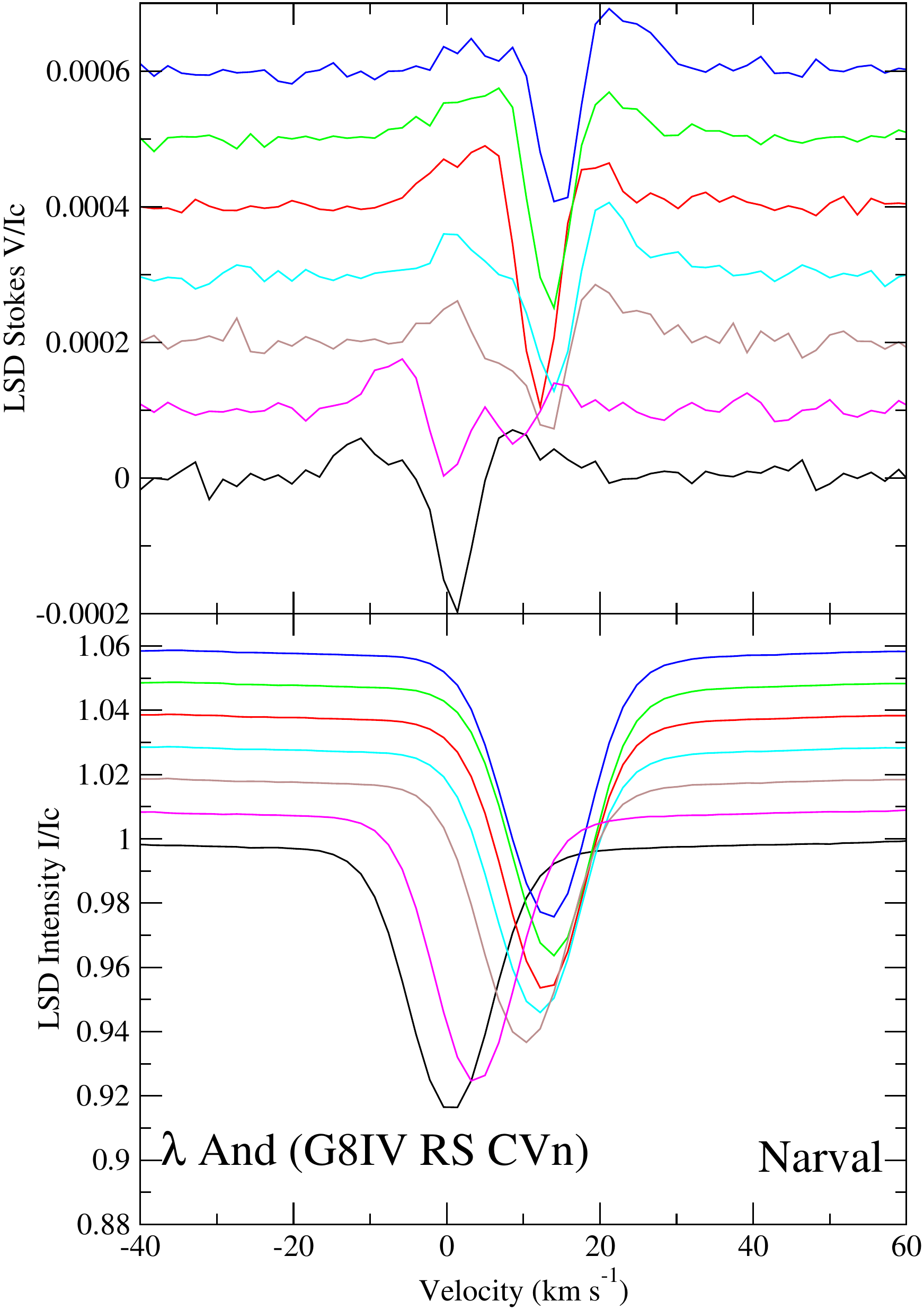}      
  \caption{Examples of magnetic field detections in the cool stars $\iota$\,Peg
and $\lambda$\,And, that have been follow-up with several observations.
$\iota$\,Peg is a binary star: the magnetic signature follows the radial
velocity of the magnetic component. The panels are the same as in Figs.~\ref{singlecool}.}
  \label{bincool}
\end{figure}

\section{The BRITE spectropolarimetric survey}

There are $\sim$600 stars brighter than V=4, which are the prime targets of
BRITE. We started a systematic survey of all these BRITE targets with
spectropolarimetry. Narval at TBL is used for all targets with declination above
-20$^\circ$, ESPaDOnS at CFHT for stars with declination between -45 and
-20$^\circ$, and HarpsPol at ESO for stars with declination below -45$^\circ$.

From the results of the MiMeS project \citep{wade2014}, we know that $\sim$10\%
of all O and B stars have detectable magnetic fields. A similar occurrence is
found for A stars and down to F5. The magnetic fields observed in these stars
are stable oblique dipoles of fossil origin, with surface strength at the poles
from $B_{\rm pol} \simeq 100$ G to several kG. Therefore we aim at detecting all
fields above 50 G. For stars cooler than F5, the magnetic fields have a dynamo
origin and $\sim$50\% of them are found to be magnetic on average
\citep[see][]{konstantinova2014}. The cool giants and supergiants, however, have
very weak fields with $B_{\rm pol}$ of the order of a few to 10 G. Therefore for
these stars, we aim at detecting all fields above $B_{\rm pol}$ = 5 G.  For each
star, we thus acquire one observation with a very high signal-to-noise, to reach
the desired detection level.

Thanks to this very high signal-to-noise spectropolarimetric observation of
each target, we will:

(1) discover new magnetic stars. This is particularly crucial for massive stars,
since only $\sim$65 magnetic OB stars are known as of today, including only a
handful of pulsating massive stars \citep[see][]{petit2013}. Note that one
measurement is enough to detect a field as magnetic  signatures appear in Stokes
V profiles even for cross-over phases (i.e. when the longitudinal field is
null).

(2) help select the best high priority targets for BRITE, i.e. the magnetic
massive ones and the most interesting cool ones. BRITE can observe all $\sim$600
stars in 6 years if each field is observed 3 months on average, but it is useful
to oberve the most interesting targets first or longer. In particular the BRITE
sample includes 11 O stars, 160 B stars (including 29 known $\beta$\,Cep stars,
20 known classical Be stars, and 22 chemically peculiar B stars), 106 A stars
(including 6 known Ap stars), 12 eclipsing binaries, 7 known $\delta$\,Scuti
stars, 7 HgMn stars, 3 RR Lyrae stars, 1 known roAp, 22 cool sub-giant stars,
several dozens red giants,... Magnetic stars among them will be prime targets
for asteroseismology.

(3) determine the fundamental parameters of all targets for the BRITE seismic
modelling: effective temperature, gravity, projected rotation velocity (vsini),
as well as abundances in particular for magnetic and chemically peculiar stars
(HgMn, Ap, Am...). See e.g. \cite{fossati2014}.

(4) provide a complete spectropolarimetric census of bright (V$\le$4) stars, by
combining the Narval, ESPaDOnS and HarpsPol data, as well as archival data.

\section{First results}

\begin{figure}[ht!]
 \centering
 \includegraphics[width=0.36\textwidth,clip]{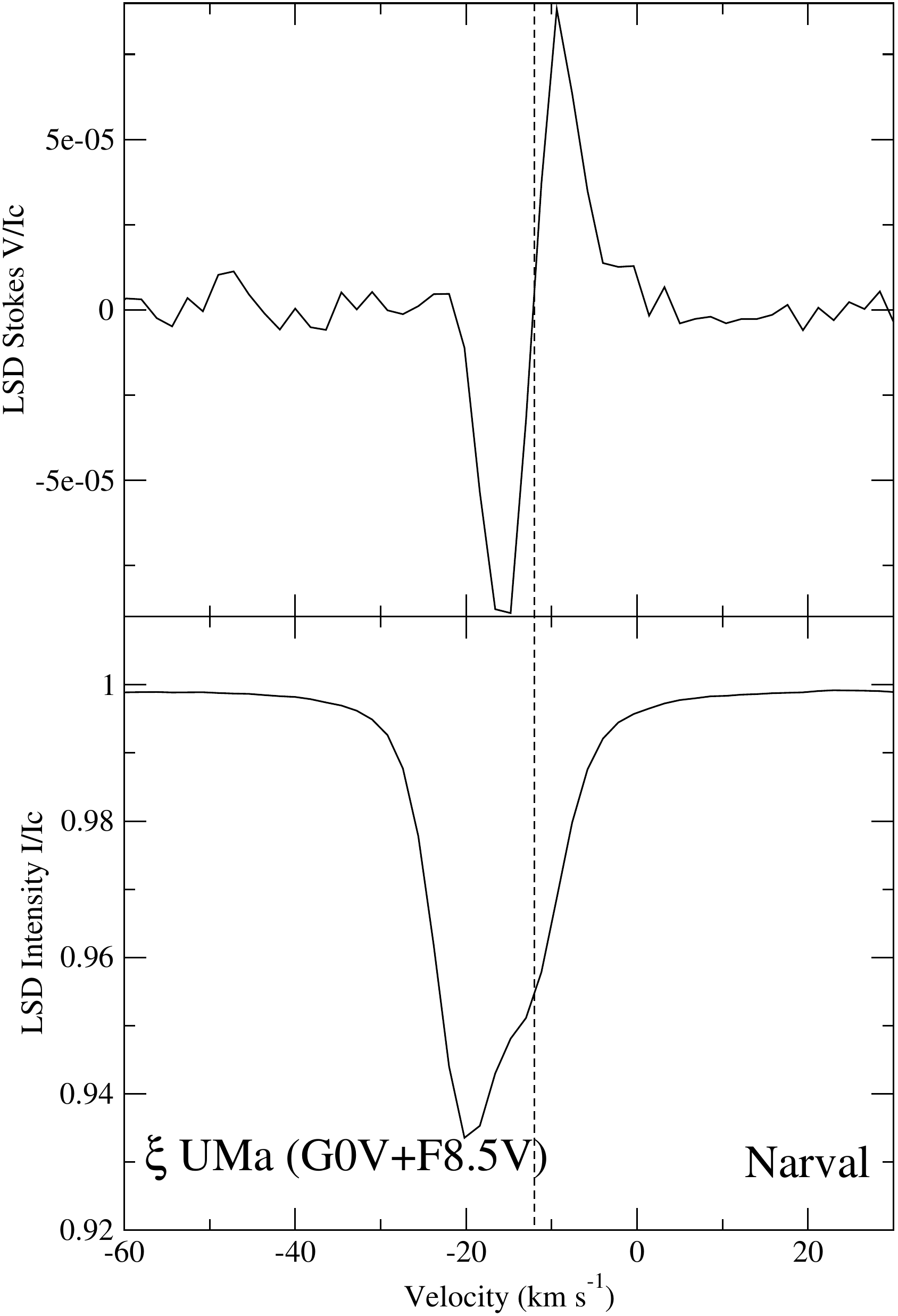} 
  \caption{Example of a magnetic field detection in the cool binary $\xi$\,UMa
observed only one time. The field can still be attributed to the red-shifted
component. The panels are the same as in Figs.~\ref{singlecool}.}
  \label{onetimebin}
\end{figure}

The first Narval and ESPaDOnS observations already led to the discovery of 14
new magnetic stars. All of them are cool stars and several are binary objects.
Examples of new detections obtained with Narval are shown in Figs.~\ref{singlecool}, \ref{bincool} and \ref{onetimebin}.

\section{Conclusions}

By combining the data acquired with Narval, ESPaDOnS and HarpsPol, a complete
spectropolarimetric census of bright (V$\le$4) stars will be available. We will
use this database to perform detailed unbiased statistics on the presence of
detectable magnetic fields in stars. The data will also be made available to the
community as a legacy, through the PolarBase database \citep{petit2014}.

All the spectra (whether the star is magnetic or not) will also serve to
determine the fundamental parameters of the BRITE stars, needed for seismic
modelling. For magnetic stars, chemical peculiarities may appear in the spectra
and will be studied as well. The best targets will be followed-up to
characterise their magnetic fields in details and provide crucial inputs for
seismic modelling.

\begin{acknowledgements}
This project is based on observations obtained at the Telescope Bernard Lyot
(USR5026) operated by the Observatoire Midi-Pyr\'en\'ees, Universit\'e de
Toulouse (Paul Sabatier), Centre National de la Recherche Scientifique (CNRS) of
France, at the Canada-France-Hawaii Telescope (CFHT) operated by the National
Research Council of Canada, the Institut National des Sciences de l'Univers of
the CNRS of France, and the University of Hawaii, and at the European Southern
Observatory (ESO), Chile.
\end{acknowledgements}

\bibliographystyle{aa}  
\bibliography{Neiner2} 

\end{document}